\documentclass{MIPRO}
\usepackage{textcomp}

\usepackage{cite}
\usepackage{amsmath,amssymb,amsfonts}
\usepackage{algorithmic}
\usepackage{graphicx}
\usepackage{textcomp}
\usepackage{xcolor}
\usepackage[T1]{fontenc}
\usepackage{flushend}
\usepackage{multirow}
\usepackage{multicol}
\usepackage{array}
\usepackage{url}

\begin{document}

\title{AI-Assisted Unit Test Writing and Test-Driven Code Refactoring: A Case Study}

\author{
\IEEEauthorblockN{
Ema Smolić\IEEEauthorrefmark{1},
Mario Brčić\IEEEauthorrefmark{1}\IEEEauthorrefmark{2},
Luka Hobor\IEEEauthorrefmark{2},
Mihael Kovač\IEEEauthorrefmark{2}
}

\IEEEauthorblockA{\IEEEauthorrefmark{1} %1st affiliation 
It from Bit d.o.o., Zagreb, Croatia}

\IEEEauthorblockA{\IEEEauthorrefmark{2} %2nd affiliation 
Faculty of Electrical Engineering and Computing, University of Zagreb, Croatia}

mario.brcic@fer.hr

}

\maketitle

\begin{abstract}
Many software systems originate as prototypes or minimum viable products (MVPs), developed with an emphasis on delivery speed and responsiveness to changing requirements rather than long-term code maintainability. While effective for rapid delivery, this approach can result in codebases that are difficult to modify, presenting a significant opportunity cost in the era of AI-assisted or even AI-led programming. In this paper, we present a case study of using coding models for automated unit test generation and subsequent safe refactoring, with proposed code changes validated by passing tests. The study examines best practices for iteratively generating tests to capture existing system behavior, followed by model-assisted refactoring under developer supervision. We describe how this workflow constrained refactoring changes, the errors and limitations observed in both phases, the efficiency gains achieved, when manual intervention was necessary, and how we addressed the weak value misalignment we observed in models. Using this approach, we generated nearly 16,000 lines of reliable unit tests in hours rather than weeks, achieved up to 78\% branch coverage in critical modules, and significantly reduced regression risk during large-scale refactoring. These results illustrate software engineering's shift toward an empirical science, emphasizing data collection and constraining mechanisms that support fast, safe iteration. 
\end{abstract}

\renewcommand\IEEEkeywordsname{Keywords}

\section{Introduction}
Many software systems begin as prototypes or minimum viable products (MVPs), which are optimized for speed rather than long-term maintainability. Shifting requirements and rapid feature integration often result in codebases that are less than optimal in terms of structure and best-practice coding. As these codebases mature, technical debt may turn small changes into complex problems, hindering further development. This issue becomes even more pronounced in the age of AI-assisted (or even AI-led) programming. Since coding models perform best in predictable, previously seen environments, poorly organized code limits their effectiveness. That presents an opportunity cost \cite{Shethiya_2025} which must be mitigated through refactoring. If refactoring is the fastest path to maintainability, then tests are the safety mechanism that makes it feasible, especially when code changes are automated. 

Tests serve as executable specifications and guardrails, limiting performance regressions and bug introductions during code refactoring \cite{fowler2018refactoring}. Research shows that LLMs can write tests more effectively than other automated test generation solutions \cite{10329992}. While studies outline some issues such as coverage ceilings and brittleness, they also propose solutions through relatively simple techniques such as better prompting strategies \cite{11029762, 10.1145/3643769, 10.1145/3663529.3663801, MUNLEY20241}.

Likewise, LLMs can refactor code effectively, leading to measurable improvements. They have been shown to reduce complexity and line count \cite{10479398, 10.1007/978-3-032-12089-2_26}. However, LLMs are also prone to hallucinations, which may result in error introduction. To address this issue, researchers propose human-in-the-loop workflows \cite{11052685,liu2024empiricalstudypotentialllms}. In this work, we explore whether automated tests can serve this purpose instead.

This experiment is motivated by a production, real-world system that originated as an MVP. To facilitate AI-based development, it needed a refactor. The system lacked a comprehensive test suite, making such refactors risky. Since refactoring by human developers would be too costly, we chose to use LLMs. We wanted to explore the feasibility of a single integrated workflow where LLMs (1) build a comprehensive, behavior-capturing test suite, and then (2) use that suite to constrain and validate model-assisted refactoring in a mature commercial codebase. 

In this paper, we present a case study evaluating whether AI-assisted unit test generation can efficiently capture existing behavior well enough to enable safe, large-scale refactoring, also led by a model. We find that it is indeed feasible. We describe the workflow in detail to aid replication in other systems and to inform best practices for similar approaches. Common failure modes in both phases are documented, and outcomes are quantified alongside qualitative analysis.

The rest of the paper is organized as follows. Section \ref{sec:rel_work} reviews prior work on code refactoring and testing, with a focus on AI-based approaches. Section \ref{sec:methodology} describes our workflow in detail. Sections \ref{sec:results} and \ref{sec:discussion} present quantitative results and discuss the findings. Section \ref{sec:threats} addresses threats to validity, and Section \ref{sec:conclusion} concludes with directions for future work.

\section{Related Work} \label{sec:rel_work}

\subsection{Refactoring and testing foundations}

Tests in software development are considered not only a means of assurance that code behaves as expected, but also an enabler of change. They serve as guardrails for code modification, with passing indicating the system remains aligned with the specifications. Similarly, tests have been described as \textit{executable specifications}, bridging the concepts of static documentation/specification and dynamic, real-time feedback on code status. Due to these benefits, they are widely used in code refactoring to ensure that modifications do not break current behavior \cite{fowler2018refactoring}.

Common concerns during the refactoring process include introducing bugs and causing regressions. Research shows that more than half of bug-inducing commits also contain refactoring changes \cite{bagheri2022refactoring}. Similarly, performance regression is widespread throughout different code change scenarios \cite{8094434}.

\subsection{AI/LLM-Assisted Testing}

Recent advances in LLMs (large language models) have prompted growing research on their application in software development, including automatic test writing \cite{10172763,10329992,11029762, 10.1145/3643769, 10.1145/3663529.3663801, MUNLEY20241}.

LLM-generated tests consistently achieve superior statement and branch coverage when compared to other automated techniques \cite{10329992}.

However, LLM-generated tests still show relatively low coverage overall. Other common concerns are their insufficient mocking capabilities and overall brittleness. However, research shows that these problems can be addressed with relatively simple techniques such as better prompting \cite{11029762, 10.1145/3643769, 10.1145/3663529.3663801, MUNLEY20241}.

\subsection{AI/LLM-Assisted Refactoring}

Similarly, LLM applications have been widely studied for automatic code refactoring \cite{11052685, cordeiro2024empiricalstudycoderefactoring}. Approaches range from the development of custom models \cite{polu2025ai} to specialized RAG techniques \cite{xu2025mantraenhancingautomatedmethodlevel}, with models used either in isolation or as subdomain experts in multi-agent systems \cite{10971563, xu2025mantraenhancingautomatedmethodlevel}.

LLMs show clear benefits in code refactoring. The generated code consistently has lower complexity and line count than the original source \cite{10479398, 10.1007/978-3-032-12089-2_26}. Moreover, even general models (i. e., not trained for code only) successfully improve code in less than 10 retries \cite{10479398} most of the time.

A known issue with LLM-based refactoring is hallucination and error introduction \cite{11052685,liu2024empiricalstudypotentialllms}. Researchers suggest that this limits their reliability in automatic pipelines and recommend having a human developer in the loop \cite{11052685}. Furthermore, LLMs perform rather poorly at identifying refactoring opportunities, but can be improved with specialized prompting techniques \cite{liu2024empiricalstudypotentialllms}.

\section{Methodology} \label{sec:methodology}

\subsection{Context}
The AI-driven testing and refactoring process discussed in this case study was applied to the frontend codebase of a commercial web application. It was developed as a React/Next.js application comprising approximately 19k LOC (lines of code), mostly TypeScript/TSX ($\sim$17.5k LOC). It uses patterns such as route groups, nested layouts, hooks, shared components, API clients, and an external component library. No consistent testing had been implemented previously.

\subsection{Models and Tooling} \label{subsec:models}

The workflow employed a hierarchical, multi-agent approach to balance reasoning depth with execution speed. The "planner" model was Gemini 2.5 Pro, accessed via the Gemini CLI. This model was selected for its large context window, which allowed it to process the entire frontend codebase at once to identify cross-module dependencies and plan refactoring steps. The "executor" models, responsible for repetitive test implementation and localized refactoring, consisted of a combination of Cursor's integrated models in Auto mode.

The prompting strategy followed a structured "Plan-Act-Verify" loop. The planner generated high-level markdown instructions based on the codebase context. These plans, along with persistent project-specific rule files (e.g., \texttt{GEMINI.md}, \texttt{.cursorrules}), served as the system prompt for the executor models. This approach ensured that even when an executor model operated on a narrow, file-specific scope, it remained constrained by global architectural rules and naming conventions.
The core architectural constraints and generative guidelines were captured in persistent, version-controlled rule and planning files, whose structure is described below.

\subsection{Stage 1: AI-Assisted Test Suite Construction} \label{subsec:methodology_test}

\begin{figure}
    \label{fig:test_flow}
    \centering
    \includegraphics[width=0.9\linewidth]{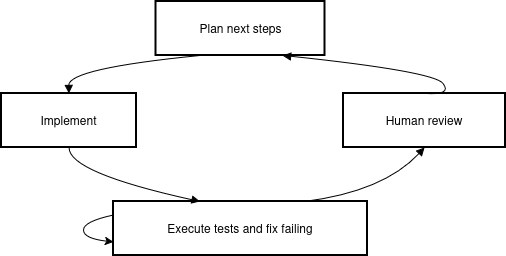}
    \caption{AI-led test generation flow}
\end{figure}

In this subsection, we outline the workflow for AI-assisted test suite construction. First, we provide guidance on specifying ground rules. Then, we define a plan-led and self-correcting incremental test generation process involving a human-in-the-loop (illustration in figure \ref{fig:test_flow}).

Prior work shows that well-formed instructions lead to better output \cite{torka2024optimizingaiassistedcodegeneration,Taeb2024}. Furthermore, research indicates that AI-based code generation without rigorous specification can introduce vulnerabilities \cite{Taeb2024}. Therefore, we employ an AI-led coding approach that relies heavily on rules and context statements. One can use both model-specific file names, such as \texttt{CLAUDE.md} \cite{anthropic_claude-code-best-practices}, which may require a special format, and broader specification files shared between agents, such as \texttt{AGENTS.md} \cite{agents-md}. Custom rule files can also be used. In these files, we captured naming conventions, import rules, and broader test-generation guidelines. Lastly, testing principles, such as prohibitions on mocking internal hooks and on modifying source code, were established.

Generating code using AI models is more effective when guided by a plan, either through self-planning \cite{jiang2024self} or using a stronger or more specialized model as the planner \cite{xie2025empoweringaigeneratebetter}. Inspired by research, we decided to use a stronger model for the test planning step while allowing a smaller, cheaper coder to handle the coding.

The plan lived in a markdown file and contained information on
\begin{enumerate}
    \item which parts of the codebase should be tested in the next step,
    \item how specific test suites should be structured, and
    \item which conventions to follow (alongside the rules in context files).
\end{enumerate}
It was updated iteratively after the previous step was implemented and reviewed by a human reviewer.

After devising a plan for a particular iteration, a coding model implemented the steps. A human did not review the code generated during development. Instead, quality was assured by the tests passing when run on existing source code. In case of failures, a limited number of fixing retries was permitted before eliminating the failing tests. However, after a completed generation (including retries) and mutation testing that eliminated ineffective tests, a developer reviewed the entire iteration, inspected failure logs, modified code as needed, and approved or rejected the changes. In Amdahl's terms, we shrank the serial part of the workflow to a thin slice, letting automation carry the parallelizable bulk.

\subsection{Stage 2: AI-Assisted Test-Guarded Refactor} \label{subsec:methodology_refactor}

\begin{figure}
    \label{fig:refactor_flow}
    \centering
    \includegraphics[width=0.9\linewidth]{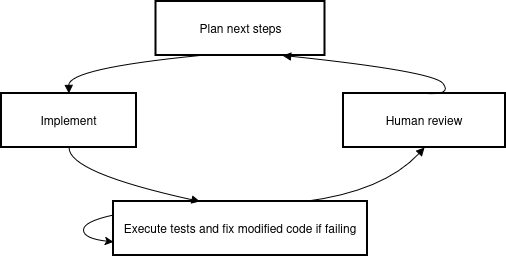}
    \caption{AI-led refactoring flow}
\end{figure}

After generating a complete test suite, the next stage in our workflow is a test-guarded refactor. The idea was to employ a coding model to refactor specific parts of the codebase in line with software development best practices (e.g., increasing modularity, reducing source file sizes, and alleviating complexity hotspots). Rules, including coding guidelines and restrictions, were contained in a special rule file (see previous section). Most importantly, test files could not be modified, except in rare cases, such as when variables were renamed in refactored code. For an illustration of this workflow, refer to figure \ref{fig:refactor_flow}. Similarly to the previous stage, the core artifacts were the mentioned rule files and a planning file shared across iterations. 

\section{Results} \label{sec:results}

In this section, we report on quantitative and qualitative (structural) changes introduced in both steps detailed in the previous section.

\subsection{AI-Assisted Test Suite Construction}

\paragraph{Generated code metrics}
During the experiment, the tests expanded from a limited baseline focused on select E2E (end-to-end) scenarios to a suite covering most frontend features. Specifically, the state after construction is:
\begin{enumerate}
    \item 87 test files containing 382 individual test cases
    \item $\sim$11,000 LOC counting only test specification
    \item $>$16,000 LOC including mocks, fixtures, and setup utilities
    \item 78.12\% branch coverage and 67.85\% line coverage in targeted, logic-heavy modules \footnote{Excludes areas like \texttt{i18n}, themes and hooks.}
\end{enumerate}

\paragraph{Test suite organization}
The test suite shows a shift from a narrow, flow-based testing approach to one that is modular, systematic, and organized around architectural boundaries. Test files were grouped by the areas they cover, which also suggested a natural order for the refactoring step. The breakdown by the number of covered cases, files, and LOC (counting only specifications) is shown in Table \ref{tab:test_struct}.

\begin{table*}[ht]
\centering
\caption{Distribution of Test Suite by Architectural Area (Ordered by Test Cases)}
\label{tab:test_struct}
\begin{tabular}{lrrr}
\hline
\textbf{Area} & \textbf{Test Cases} & \textbf{Spec Files} & \textbf{Spec LOC} \\
\hline
Reusable UI components   & 127 (33.2\%) & 32 (36.8\%) & 2,801 (25.0\%) \\
Feature-specific modules & 121 (31.7\%) & 23 (26.4\%) & 5,010 (44.7\%) \\
Application pages        & 57 (14.9\%)  & 17 (19.5\%) & 1,702 (15.2\%) \\
API-facing logic         & 55 (14.4\%)  & 7 (8.0\%)   & 1,030 (9.2\%)  \\
Authentication           & 18 (4.7\%)   & 7 (8.0\%)   & 582 (5.2\%)    \\
Utilities                & 4 (1.0\%)    & 1 (1.1\%)   & 74 (0.7\%)     \\
\hline
\textbf{Total}           & 382 (100\%)  & 87 (100\%)  & 11,199 (100\%) \\
\hline
\end{tabular}
\label{tab:test_distribution}
\end{table*}

\paragraph{Tests as documentation}
Test-to-source parity increased from a negligible baseline ratio to one approaching 1:1 with TypeScript/TSX ($\sim$17.5k LOC). That makes the test suite useful as documentation of expected behavior and justifies its use as a guard during the automated refactoring step.

\subsection{AI-Assisted Test-Guarded Refactor}

\paragraph{Code evolution metrics}
Contrary to the previous step, where the aim was to increase the amount of test code, the goal in the refactoring step was more complex. While it is generally desirable to have fewer lines of source code for readability and maintainability, reducing the number of lines must not result in lower-quality code organization. Following the experiment, we found that:
\begin{itemize}
    \item 219 files modified (146 additions, 120 deletions, 51 in-place modifications)
    \item source file count grew from 237 to 263 (+26 files)
    \item LOC increased from 18,619 to 21,624 (+3,005 LOC)
\end{itemize}
As we show below, this growth reflects logical redistribution rather than bloat. The number of deleted files nearly matches the number of additions (120 vs.\ 146), already suggesting that the diffs are driven by architectural restructuring.

\begin{figure}[h]
    \label{fig:post_refactor_source}
    \centering
    \includegraphics[width=0.9\linewidth]{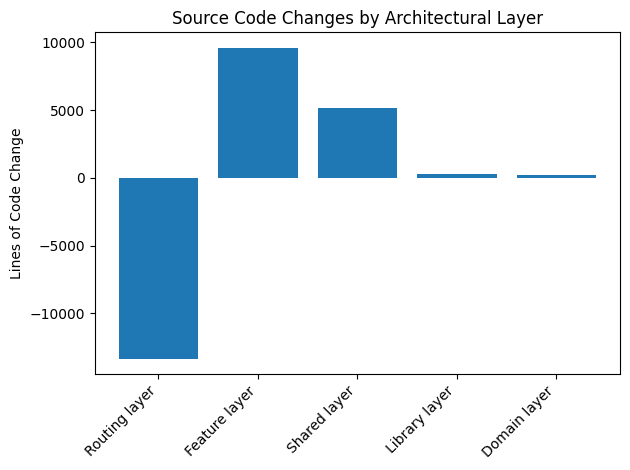}
    \caption{LOC difference pre- and post-refactor}
\end{figure}

\paragraph{Source code distribution}
Before the refactor, most logic was concentrated in the routing layer, with no dedicated feature or shared layers. After the refactor, the routing layer became much leaner and new layers were introduced. The breakdown of changes by layers is shown in Table \ref{tab:layer_distribution_textloc} and illustrated in Figure \ref{fig:post_refactor_source}. That supports the claim that the increase in files and LOC results from architectural restructuring rather than code bloat.

\begin{table*}[h]
\centering
\caption{Distribution of LOC in \texttt{frontend/src} by Architectural Layer (Before $\to$ After Refactor)}
\label{tab:layer_distribution_textloc}
\begin{tabular}{lrrrr}
\hline
\textbf{Layer} & \textbf{Text files ($\Delta$)} & \textbf{Text LOC ($\Delta$)} & \textbf{Share of Text LOC} & \textbf{Avg LOC/File} \\
\hline
Routing (src/app)
& 166 $\to$ 52 ($\Delta$ = $-$114)
& 17,872 $\to$ 6,201 ($\Delta$ = $-$11,671, $-$65.3\%)
& 96.0\% $\to$ 28.7\%
& 108 $\to$ 119 \\

Features (src/features)
& 0 $\to$ 79 ($\Delta$ = +79)
& 0 $\to$ 9,595 ($\Delta$ = +9,595)
& 0\% $\to$ 44.4\%
& 0 $\to$ 121 \\

Shared (src/shared)
& 0 $\to$ 63 ($\Delta$ = +63)
& 0 $\to$ 5,130 ($\Delta$ = +5,130)
& 0\% $\to$ 23.7\%
& 0 $\to$ 81 \\

Domains (src/domains)
& 0 $\to$ 3 ($\Delta$ = +3)
& 0 $\to$ 198 ($\Delta$ = +198)
& 0\% $\to$ 0.9\%
& 0 $\to$ 66 \\

Libraries (src/libs)
& 13 $\to$ 8 ($\Delta$ = $-$5)
& 532 $\to$ 283 ($\Delta$ = $-$249, $-$46.8\%)
& 2.9\% $\to$ 1.3\%
& 41 $\to$ 35 \\

Other text (assets + misc)
& 47 $\to$ 47 ($\Delta$ = 0)
& 215 $\to$ 217 ($\Delta$ = +2, +0.9\%)
& 1.2\% $\to$ 1.0\%
& 5 $\to$ 5 \\

\hline
\textbf{Total \texttt{frontend/src} (text only)}
& 226 $\to$ 252 ($\Delta$ = +26)
& 18,619 $\to$ 21,624 ($\Delta$ = +3,005, +16.1\%)
& 100\% $\to$ 100\%
& 82 $\to$ 86 \\

\textit{TS/TSX text (subset of total)}
& 180 $\to$ 206 ($\Delta$ = +26)
& 17,511 $\to$ 20,476 ($\Delta$ = +2,965, +16.9\%)
& 94.0\% $\to$ 94.7\%
& 97 $\to$ 99 \\

\textit{Non-TS/TSX text (subset of total)}
& 46 $\to$ 46 ($\Delta$ = 0)
& 1,108 $\to$ 1,148 ($\Delta$ = +40, +3.6\%)
& 6.0\% $\to$ 5.3\%
& 24 $\to$ 25 \\
\hline
\end{tabular}
\end{table*}

\section{Findings and Discussion} \label{sec:discussion}

In this section, we discuss our findings based on the results stated in the previous section.

\subsection{AI-Assisted Test Suite Construction}

During the AI-led test generation process described in Section \ref{sec:methodology}, Subsection \ref{subsec:methodology_test}, we found that the model invested significant effort into shared infrastructure. In other words, tests were organized into files grouped by logical directories, and the model applied patterns such as shared setup logic, standardized mocking, and separation of test logic from test data---some of which were outlined in the rule files, but others emerged on their own as best-practice \textit{common sense}.

For instance, the model introduced a centralized mock data catalog of several thousand LOC with consolidated fixtures, reducing duplication. It also established an API-mocking layer based on request interception and a global test harness with consistent environment setup, browser-like polyfills, and strict side-effect handling. This infrastructure enabled reuse and further growth.

\paragraph{Adaptive code organization through iterations}
Also, we noticed that test code was iteratively refactored, consolidated, modified, and removed (to reduce brittleness or redundancy). That was allowed in the setup and shows that coding models can recognize issues and improve their own code. Moreover, conventions and infrastructure evolved alongside new tests, necessitating occasional refactoring. LLMs work best with deterministic feedback, be it automated or human-given. We found that LLMs left to their own devices produced a high proportion of unit tests (approaching 40\%) that contribute to code coverage (as stated as an objective) but are ineffective. That is a form of \textit{value misalignment} in which lazy but performant solutions are chosen, unless the qualities are meticulously stated and enforced across other important dimensions. Having a \textbf{domain-specific model harness} is highly important; in this case, for the test generation phase, we used mutation testing to identify ineffective tests that could be eliminated or changed. In total, we find that allowing already generated, human-approved code to evolve over iterations results in a more internally consistent suite. 

\subsection{AI-Assisted Test-Guarded Refactor}

\paragraph{Structurally invasive refactor}
Regarding the refactoring step, we noticed that the changes significantly altered the structure while preserving test-guarded behavior. It contained a major redistribution of logic into specialized layers, with only a modest growth in LOC. The result is an architecture based on reusable components and clear layer boundaries.

\paragraph{On LOC growth and architectural interpretation}
The observed 16.1\% growth in total LOC may appear counterproductive in isolation. However, several structural indicators suggest genuine architectural improvement rather than simple code redistribution. First, the routing layer shrank by 65.3\% in LOC, shifting from containing 96\% of all code to 28.7\%. Second, this logic was distributed across three new layers (features, shared, domains), each with a distinct architectural role. Third, the average LOC per file remained stable across the codebase (82 to 86), indicating that the new files are comparable in granularity to existing ones rather than aggregating unrelated functionality. Fourth, 120 files were deleted alongside 146 additions, indicating that the changes involved restructuring rather than solely additive growth.

To move beyond LOC-level observations, we parsed the codebase's Abstract Syntax Tree (AST) to extract finer-grained structural metrics before and after the refactor.

\textbf{Coupling (Dependency Analysis).} In the pre-refactor state, the routing layer (\texttt{src/app}) acted as a monolithic container heavily coupled to business logic, evidenced by 893 unique internal \texttt{import} statements. Following the refactor, internal imports within the routing layer were reduced to 379 — a 57.5\% reduction in dependency density. The newly created \texttt{features} and \texttt{shared} layers absorbed 467 and 186 modular imports, respectively, confirming a shift toward higher functional cohesion.

\textbf{Cyclomatic Complexity.} While the total number of functions and components grew from 806 to 1,022 to support the new modular architecture, the average cyclomatic complexity per function decreased from 2.24 to 2.13. Within the routing layer specifically, average function complexity dropped to 1.97.

\textbf{Modularization vs. Aggregation.} The increase in average LOC per file noted in Table \ref{tab:layer_distribution_textloc} (e.g., 108 to 119 in the routing layer, 0 to 121 in features) reflects deliberate modularization rather than code bloat. In the new \texttt{features} layer, files contain an average of 6.49 cohesive functions, with a low average complexity of 2.19 per function, indicating that previously scattered logic has been grouped into cohesive, self-contained modules.

\section{Threats to Validity} \label{sec:threats}

\paragraph{Internal validity}
The main threat to internal validity is residual value misalignment that we have not yet identified. The methodology's reliance on persistent configuration files (\texttt{GEMINI.md}, \texttt{.cursorrules}) and versioned Markdown plans ensures that the core generative constraints are documented. The test suite was validated by passing against existing source code and by surviving selection pressure in the model harness, aligning the work with software engineering's best practices. However, it is quite possible that codebase quality did not improve or even degraded on some important aspects that we did not specify or measure in the harness or objectives. We iteratively extended our criteria and metrics considered relative to the starting point throughout the procedure. 

\paragraph{External validity}
This study examines a single commercial frontend codebase built with React/Next.js \cite{yin2018case, runeson2009guidelines}. The findings may not transfer to other technology stacks, backend systems, or codebases of different scales or maturity levels. Replication across diverse settings — particularly traditional object-oriented backends and legacy systems — is needed to establish broader applicability.

\paragraph{Construct validity}
Refactoring quality was evaluated using both LOC-level distributions and AST-based metrics (cyclomatic complexity and dependency analysis). These are well-suited to the functional, component-based React paradigm studied here. Test quality was assessed through branch and line coverage, following constraint satisfaction on test effectiveness.

\section{Conclusion and Further Research} \label{sec:conclusion}

To sum up, the first step of the experiment enriched an undertested codebase with a large, structured, infrastructure-backed unit test suite. It introduced hundreds of test cases across core architectural areas. The generated test suite followed best practices, especially through substantial shared infrastructure that supports reuse and stability. 

On the other hand, the refactor step transformed a poorly-organized MVP into a layered architecture with improved modularity. While the overall source code size grew moderately, the dominant effect was a redistribution of logic from routing files into dedicated feature and shared layers. The refactor involved extensive file relocation and selective rewriting, yet it preserved externally observable behavior. AST-based analysis confirms that the resulting structure reduces coupling (57.5\% fewer imports in the routing layer), lowers average cyclomatic complexity (2.24 to 2.13), and provides a more maintainable foundation for further development. Future work should replicate the approach across different technology stacks.
Furthermore, it should create a catalog of best practices as well as different important quality metrics and their thresholds. That should help alleviate value misalignment \cite{brcic2023impossibility} that occurs either in goal specification or specification enforcement. In this paper, we identified both issues. We addressed them through feedback and iteration during test generation (where coverage can be improved by ineffective tests that need to be pruned using mutation testing) and later code refactoring (where quality gets improved only on explicitly stated objectives). It is important to have LLM-independent, deterministic code for evaluating important quality characteristics, both as part of the model harness and in post-hoc solution analysis, as we have noticed that LLMs hesitate to report subpar metrics unless explicitly pressed on the issue, while pushing positive metrics and aspects to the forefront. Such model behavior is consistent with the recent research \cite{sofroniew2026emotion}.

In conclusion, these two phases demonstrate that AI-assisted workflows can be effectively applied to mature, production systems, provided that initial guidance is set in rule files, model harness is in place, and a human-in-the-loop is present at most critical moments. By combining automated test generation with automated refactoring, the approach builds on current advances in AI-assisted coding.

\bibliographystyle{IEEEtran}
\bibliography{references}

\end{document}